\documentclass[fleqn,usenatbib]{mnras}

\usepackage{newtxtext,newtxmath}

\usepackage[T1]{fontenc}

\DeclareRobustCommand{\VAN}[3]{#2}
\let\VANthebibliography\thebibliography
\def\thebibliography{\DeclareRobustCommand{\VAN}[3]{##3}\VANthebibliography}

\usepackage{graphicx}	
\usepackage{amsmath}	

\usepackage{booktabs}
\usepackage{hyperref}
\usepackage{mathrsfs}
\usepackage{natbib}
\usepackage{appendix}
\usepackage{xcolor}
\usepackage{subcaption}

\title[Neutrino Observations of LHAASO Sources]{Neutrino Observations of LHAASO Sources: Present Constraints and Future Prospects}

\author[Huang and Li]{
Tian-Qi Huang,$^{1,2}$\thanks{E-mail: htq@pku.edu.cn}
and Zhuo Li,$^{1,2}$\thanks{E-mail: zhuo.li@pku.edu.cn}
\\
$^{1}$Department of Astronomy, School of Physics, Peking University, Beijing 100871, People's Republic of China\\
$^{2}$Kavli Institute for Astronomy and Astrophysics, Peking University, Beijing 100871, People's Republic of China\\
}

\date{Accepted XXX. Received YYY; in original form ZZZ}

\pubyear{2021}

\begin{document}
\label{firstpage}
\pagerange{\pageref{firstpage}--\pageref{lastpage}}
\maketitle

\begin{abstract}
The Large High Altitude Air Shower Observatory (LHAASO) observed a dozen of gamma-ray sources with significant emission above 100 TeV, which may be strong candidates of PeVatrons. Neutrino observations are crucial to diagnose whether the gamma-ray radiative process is hadronic or leptonic. We use the Bayesian method to analyze the ten-year (2008-2018) IceCube data, and hence constrain the hadronic gamma-ray emission in the LHAASO sources. The present neutrino data show that the hadronic gamma-ray flux from Crab Nebula  is lower than the observed gamma-ray flux at the 90\% C.L. and contributes less than 86\%, which disfavors the hadronic origin of the gamma-rays below tens of TeV. For the other LHAASO sources, the present neutrino observations cannot put useful constraints on the gamma-ray radiative process. 
We consider the uncertainty of the source extension: the upper limits on hadronic gamma-ray flux tend to increase with the extension; 
and some sources, i.e., LHAASO J2032+4102, LHAASO J1929+1745, and LHAASO J1908+0621, show relatively high statistical significance of neutrino signals if the extension is $\lesssim0.6^\circ$. We finally estimate the future observational results of LHAASO sources by the proposed neutrino telescopes. If the LHAASO-observed PeV gamma-rays are of hadronic origin, Crab Nebula may be detected at $>100$ TeV at $3\sigma$ C.L. within 20 years by a neutrino detector with the effective area 30 times that of IceCube.
\end{abstract}




\section{Introduction} \label{sec:intro}


The origins of high-energy cosmic-rays are still unclear. Their propagation trajectories are deflected from the direction of accelerators by the magnetic field in the medium. Thus it is difficult to identify the sources by cosmic ray observations. However, cosmic-rays interact with the background baryons or radiation in/around the source and generate high energy gamma-rays (via $\pi^{0}$ decay) and neutrinos (via $\pi^{\pm}$ decay), which are not affected by the magnetic field in propagation and can be good indicators of cosmic-ray sources. 

The most energetic photon ever observed is with 1.4 PeV and from the direction of Cygnus OB2. It was detected by the $\rm km^2$ array (KM2A) of the Large High Altitude Air Shower Observatory (LHAASO) \citep{2021Natur.594...33C}. LHAASO-KM2A had also detected significant gamma-ray emission above 100 TeV from 12 sources \citep{2021Natur.594...33C,425, 2021ApJ...919L..22C}, which are probably the Galactic accelerators of PeV cosmic-rays, namely, PeVatrons. According to gamma-ray observations, 
various PeVatron candidates had been proposed, including, e.g., the Galactic center \citep[e.g.][]{2016Natur.531..476H}, supernova remnants  \citep[SNRs, e.g.][]{2021NatAs...5..460T}, pulsar wind nebulae \citep[PWNe, e.g.][]{2012SSRv..173..341A} and young massive star clusters \citep[YMCs, e.g.][]{2019NatAs...3..561A}. 

A detection of gamma-rays up to hundreds of TeV is not sufficient to identify a source as PeVatron, because the gamma-rays can be produced by leptonic processes as well, e.g. inverse Compton scattering off electrons. However, the detection of neutrinos accompanying the $\pi^{0}$ decayed gamma-rays will definitely help to identify PeVatrons.

Most of the 12 LHAASO sources with 100 TeV emission (LHAASO sources hereafter) show TeV gamma-ray counterparts. The neutrino emission from these TeV gamma-ray counterparts have been investigated in previous works, e.g., \cite{2009NIMPA.602..117K, 2017APh....86...46H}. In some cases, sophisticated models have been proposed for the neutrino emission, for example, the Crab Nebula \citep[e.g.][]{2003A&A...402..827A}, Cygnus Region \citep[e.g.][]{2007PhRvD..75f3001A, 2021ApJ...921L..10B} and SNR G106.3+2.7 \citep[e.g.][]{2021Innov...200118G}. 

IceCube Neutrino Observatory, a $\rm km^3$ scale detector at the South Pole, has carried out searches for time-integrated neutrino signals from the TeV gamma-ray sources, but results in non-detection so far \citep[e.g.][]{2011ApJ...732...18A, 2013ApJ...779..132A, 2017ApJ...835..151A, 2019EPJC...79..234A, 2020PhRvL.124e1103A}. In the recent search using the ten-year data of IceCube from 2008 to 2018, \cite{2020PhRvL.124e1103A} searched for astrophysical neutrinos from the directions of a list of candidate sources, where MGRO J1908+06 is the most significant Galactic source but 
still far from the $3\sigma$ confidence level. The upper limits on the neutrino flux from these candidate sources are provided, under the assumptions of point-like neutrino sources and $E^{-2}$ or $E^{-3}$ neutrino spectra. Using these neutrino upper limits \cite{2022ApJ...925...85H} put constraints on the hadronic gamma-ray components from LHAASO sources.

Since the upper limits on neutrino flux depend on the assumptions of the source extensions and neutrino spectral shapes, in this work we evaluate the effect of these assumptions on the upper limit estimate. We will use the Bayesian method and the ten-year IceCube muon-track data \citep{2021arXiv210109836I} from 2008 to 2018 to search for the neutrino flux from LHAASO sources and hence constrain the hadronic components in gamma-rays. 
The current operating detectors \citep[e.g. IceCube and Baikal-GVD][]{2020arXiv201109209Z} only shows non-detection, but several neutrino telescopes are under construction \citep[e.g. KM3NeT-ARCA][]{2016JPhG...43h4001A} or being planed \citep[e.g. P-ONE and IceCube-Gen2][]{2020NatAs...4..913A,2021JPhG...48f0501A}, thus in this work we further predict the future observation results of neutrinos from the LHAASO sources.

 This paper is organized as follows. In \autoref{sec:source}, we introduce the IceCube muon-track data that is used. In \autoref{sec:method}, we describe the Bayesian method used for deriving the neutrino upper limits.  In \autoref{sec:proton}, we present the connection between the spectra of neutrinos and the associated hadronic gamma-rays. In \autoref{sec:result}, we compare the gamma-ray observations with the upper limits of hadronic gamma-ray flux and discuss the uncertainties in the upper limit estimate. In \autoref{sec:prospects}, we estimate the significance of neutrino signals from LHAASO sources in the combined search by the future projects. Finally \autoref{sec:summary} is the summary and discussion.

\begin{figure*}
\begin{center}
\includegraphics[width=0.86\textwidth]{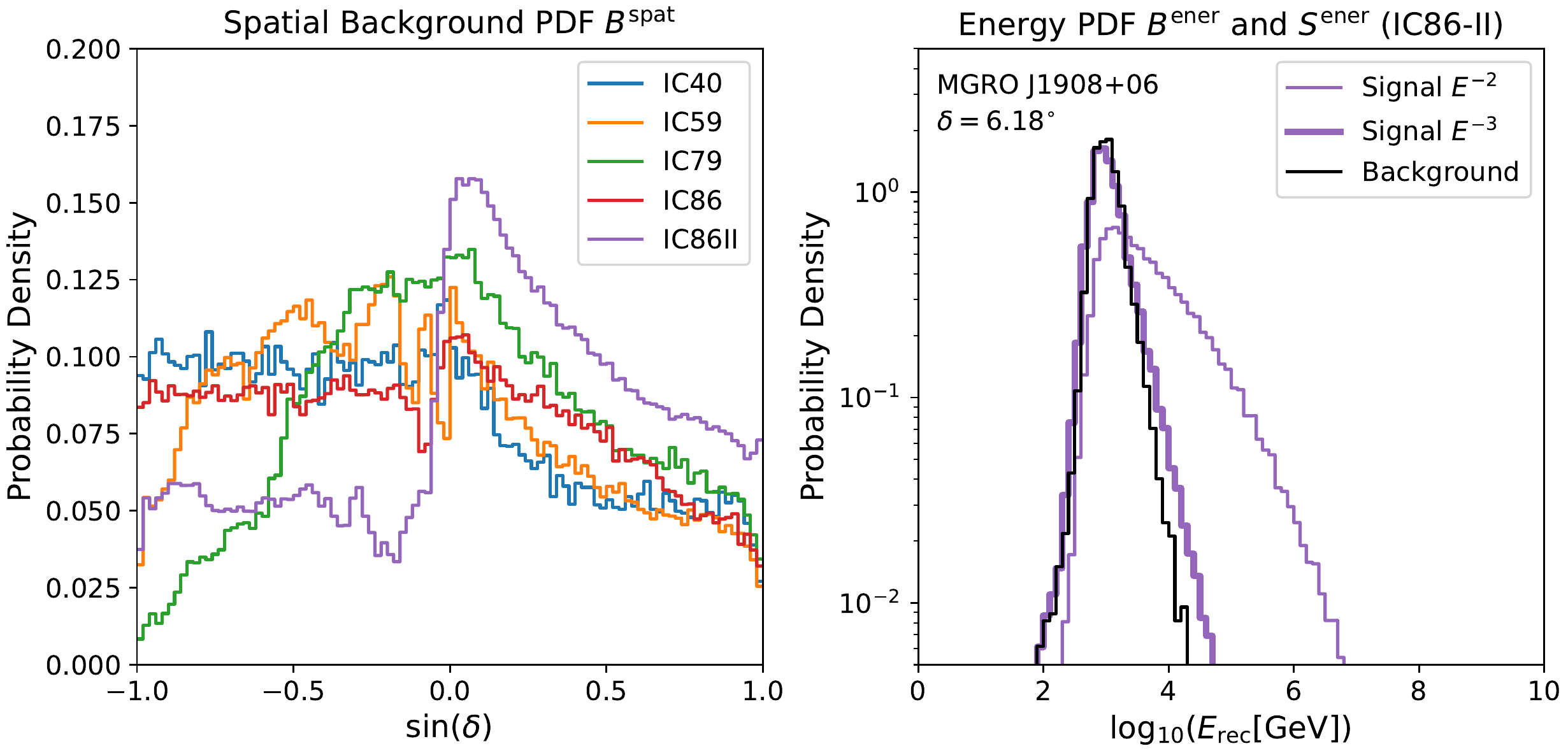}
\caption{An example of the background and signal PDFs. Left: the spatial background PDF. Different colors represent different data samples. Right: the energy background PDF and the energy signal PDF in the case of MGRO J1908+06 for data sample IC86-II. The thin (thick) purple line shows the energy signal PDF for an $E^{-2}$ ($E^{-3}$) neutrino spectrum. The black line shows the energy background PDF at $\delta_s=6.18^{\circ}$.}\label{figure_PDF}
\end{center}
\end{figure*}

\section{The muon-track data from IceCube}\label{sec:source}

IceCube released the muon-track data from 2008 to 2018 which was used in the ten-year search for point-like neutrino sources \citep{2021arXiv210109836I}. We will use these data to search for neutrino signals from the LHAASO sources. The data is composed of three parts:

\textbf{The experimental data events.} The experimental data events are grouped into five samples including IC40, IC59, IC79, IC86-I, and IC86-II, corresponding to different construction levels of the detector. The number in the sample name represents the number of strings in the detector. Digital Optical Modules on the string record the Cherenkov light from the charged particles produced in the neutrino-nucleus interactions ($\nu+N$). The interaction time, the reconstructed direction and the reconstructed energy ($E_{\rm rec}$) are given for each event. Note that $E_{\rm rec}$ is the reconstructed energy of the muon passing through the detector, different from the energy of the incident neutrino which generates the muon through the neutrino-nucleus charge current (CC) interaction $\nu_{\mu}+N{\xrightarrow{\rm CC}}\mu+X$.

\textbf{The instrument response functions.} The instrument response functions include the effective area and the smearing matrix. The effective area $A_{\rm eff}(E_{\nu},\delta_\nu)$ relies on the neutrino energy $E_{\nu}$ and the declination angle $\delta_{\nu}$. The smearing matrix $\mathcal{M}(E_{\rm rec}|E_{\nu}, \delta_{\nu})$ gives the fractional count of simulated events in the reconstructed energy bin relative to all events in the ($E_{\nu}$, $\delta_{\nu}$) bin. The matrix tells us the probability to get the reconstructed energy $E_{\rm rec}$ when a neutrino with the energy $E_{\nu}$ enters the detector from the declination $\delta_{\nu}$.

\textbf{The detector uptime.} The detector uptime records the time periods during which the detector is running well. We can get the livetime of the detector for each data sample.

\section{Bayesian method for signal search}\label{sec:method}

Since the present data only result in the non-detection of neutrino signals from LHAASO sources (see below), we should find a way to give the upper limits to the neutrino flux.
We will use the Bayesian method to search for signal events. The Bayesian method here is based on the form of unbinned likelihood function which is widely used in the neutrino source searches \citep{2008APh....29..299B, 2011ApJ...732...18A,2014ApJ...796..109A,2017ApJ...849...67A, 2019EPJC...79..234A, 2020PhRvL.124e1103A, 2021ApJ...914...91K, 2021PhRvD.103l3018Z}. Given the likelihood function one can calculate the probability density of neutrino event number, and then derive the 90\% C.L. upper limit of the neutrino flux $\Phi^{90\%}_{\nu_{\mu}+\bar{\nu}_{\mu}}$.

\subsection{Likelihood function}
The first step is to build the likelihood function. The likelihood $L$ for observing $n_s$ signal events from a source is given by the product of probability density functions (PDFs) for each track event. Consider a source at position $\vec{x}_s=(\alpha_{s}, \delta_{s})$, with $\alpha_s$ the right ascension and $\delta_s$ the declination angle, with angular extension $\sigma_{s}$. Assuming a power law neutrino spectrum $dN_{\nu}/dE_{\nu}\propto E_{\nu}^{-\gamma}$, a likelihood can be built for the source,
\begin{equation}
\label{eq:likelihood}
\begin{aligned}
L&=\prod_{k}\prod_{i}^{N_{k}}\left( \frac{n_s^k}{N_k}S_i+\frac{N_k-n^k_s}{N_k}B_i\right), \\
\end{aligned}
\end{equation}
where $S_i$ and $B_i$ are the PDFs for signal and background events, $k\in\left\{ { \rm IC40,\,IC59,\,IC79,\,\text{IC86-I},\,\text{IC86-II} } \right\}$ represents five data samples, $N_{k}$ is the total number of events in data sample $k$,  and $n^{k}_s$ is the number of signal events in data sample $k$. 

If the source extension is small, the detection probability over the entire source extension can be considered to be proportional to the effective area $A_{\rm eff}^{k}$ for the source declination $\delta_{s}$. We write
\begin{equation}
\label{eqn:(4)}
n^k_s=n_s\times\frac{\int t_k \int E_{\nu}^{-\gamma}A_{\rm eff}^{k}(E_{\nu}\,, \delta_s)dE_{\nu}}{\sum_{k^{'}} \int t_{k^{'}} \int E_{\nu}^{-\gamma}A_{\rm eff}^{k^{'}}(E_{\nu}, \delta_s)dE_{\nu}},
\end{equation} 
where $n_s$ is the total number of signal events and $t_{k}$ is the detector livetime for data sample $k$.

The PDFs for signal and background events are given, respectively, by
\begin{flalign}
\label{eq:likelihood2}
S_i&=S^{\rm spat}(\vec{x}_i|\sigma_i, \vec{x}_s, \sigma_{s})\times S^{\rm ener}(E_i|\vec{x}_s,\gamma), \\
B_i&=B^{\rm spat}(\delta_i)\times B^{\rm ener}(E_i|\delta_i).
\end{flalign}
Here $S^{\rm spat}$ is the spatial signal PDF that describes the distribution of the reconstructed direction of signal events $\vec{x}_i$. If the source follows a 2D Gaussian with extension $\sigma_{s}$, and the event angular uncertainty is $\sigma_i$, we have
\begin{equation}
S^{\rm spat}(\vec{x}_i|\sigma_i, \vec{x}_s,\sigma_{s})=\frac{1}{2\pi(\sigma_i^2+\sigma_s^2)}{\rm exp}\left[-\frac{\phi^2}{2(\sigma_{i}^2+\sigma_s^2)}\right]\frac{\phi}{{\rm sin}\phi},
\end{equation}
where $\phi$ is the angular distance between $\vec{x}_i$ and $\vec{x}_s$. The spatial signal PDF is normalized as $\int S^{\rm spat}(\vec{x}|\sigma_i,\vec{x}_s, \sigma_{s})d\Omega=1$. Typically, $\sigma_i\approx0.64^{\circ}$ \citep{2017APh....86...46H}.

Next, $S^{\rm ener}$ is the energy signal PDF that describes the distribution of the reconstructed energy of signal events, $E_{i}$,  
from the source direction $\vec{x}_s$. We use the effective area $A_{\rm eff}^{k}$ and the smearing matrix $\mathcal{M}_{k}$ to build the energy signal PDF,
\begin{equation}
\label{eqn:(6)}
S^{\rm ener}(E_i|\vec{x}_s, \gamma)=\frac{\int E_{\nu}^{-\gamma}A_{\rm eff}^{k}(E_{\nu}, \delta_s)dE_{\nu}\cdot\mathcal{M}_{k}(E_{i}|E_{\nu}, \delta_s)}{\int E_{\nu}^{-\gamma}A_{\rm eff}^{k}(E_{\nu}, \delta_s)dE_{\nu}}.
\end{equation}

Finally, $B^{\rm spat}$ denotes the spatial background PDF, and $B^{\rm ener}$ the energy background PDF. Both are nearly uniform in right ascension. We assume that the background events follow the PDF of experimental data scrambled in right ascension. For data sample $k$, we denote $N_{ij}^{k}$ the number of events with the declination ${\rm sin}\,\delta\in[{\rm sin}\delta_i,\,{\rm sin}\delta_i+0.02)$ and the reconstructed energy ${\rm log_{10}}E_{\rm rec}\in[{\rm log_{10}}E_{j}, {\rm log_{10}}E_{j}+0.1)$. Thus the spatial and energy background PDFs are given, respectively, by
\begin{flalign}
&B^{\rm spat}(\delta)=\frac{\sum_j N_{ij}^{k}}{N_{k}\times \Delta \Omega},\\
&B^{\rm ener}(E_{\rm rec}| \delta)=\frac{N_{ij}^{k}}{\sum_j N_{ij}^{k}},
\end{flalign}
where 
$\Delta \Omega=2\pi\Delta {\rm sin}\,\delta=0.04\pi$.

\autoref{figure_PDF} gives an example of the background and signal PDFs. The left panel shows the spatial background PDF in five data samples. The right panel shows the energy PDF for background and signals in the direction of MGRO J1908+06 for data sample IC86-II. MGRO J1908+06 shows the most significant neutrino signals in Galactic sources in the ten-year search \citep{2020PhRvL.124e1103A}.  

\subsection{Probability density of event number and the neutrino flux upper limit}

\begin{figure}
\begin{center}
\includegraphics[width=0.86\columnwidth]{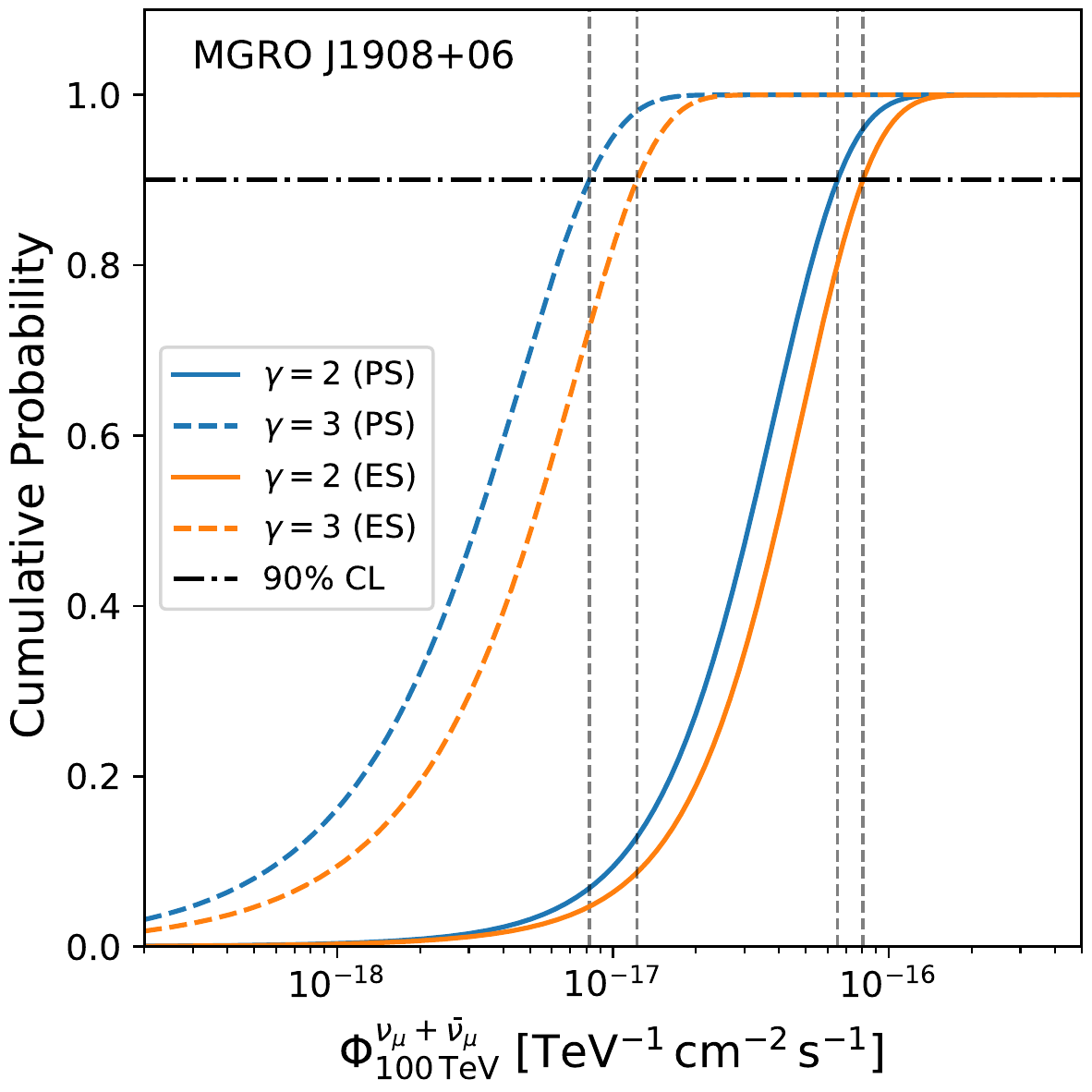}
\caption{The cumulative probability of the 100 TeV neutrino flux in the case of MGRO J1908+06. Different models correspond to different spectral index $\gamma$ and source extension, $\sigma_s=0.0^{\circ}(\rm PS)$  and $0.34^{\circ}(\rm ES)$. The horizontal line marks the 90\% confidence level. The vertical lines mark the 90\% upper limits on astrophysical neutrino flux $\Phi^{90\%}_{\nu_{\mu}+\bar{\nu}_{\mu}}$ for different models.}\label{figure_TS}
\end{center}
\end{figure}

As there is no preference on the signal event number, the prior distribution $\pi(n_s)$ is assumed to be uniform in the range $[0, N]$, where $N=\sum_{k}N_k$ is the total number of events. Given the spectral index $\gamma$ and source extension $\sigma_s$, the probability density of event number can be expressed as
\begin{equation}
p(n_s|X, \gamma, \sigma_s)=\frac{L(X|n_s, \gamma, \sigma_s)\pi(n_s)}{\int L( X|n_s, \gamma, \sigma_s)\pi(n_s) dn_s},
\end{equation}
where $X$ represents the observed data. The cumulative probability $\int^{n_{90}}p(n_s|X, \gamma, \sigma_s)dn_s=0.9$ defines the 90\% upper limit on signal event number $n_{90}$. Given $n_{90}$, the 90\% upper limit on neutrino flux $\Phi^{90\%}_{\nu_{\mu}+\bar{\nu}_{\mu}}=dN_{\nu}/dE_{\nu}$ can be derived by solving out
\begin{equation}
\label{eq03}
	n_{\rm 90}=\sum_k \int dt_k\int dE_{\rm \nu} \Phi_{\rm \nu_\mu+\bar{\nu}_\mu}^{90\%}A_{\rm eff}^{k}(E_{\rm \nu},\delta_s).
\end{equation}
For an example, \autoref{figure_TS} shows the cumulative probability function (CDF) of MGRO J1908+06 under the point source (PS) and extended source (ES) hypothesis, respectively. The extension of MGRO J1908+06 is around $0.34^{\circ}$ by \cite{2009A&A...499..723A}.

\subsection{Caveats}

Some notes should be made here. Firstly, the spectrum term $E_{\nu}^{-\gamma}$ in the likelihood function can be replaced by the other spectral shapes. So we can measure the astrophysical neutrino flux under any spectral shape assumptions.

Secondly, considering a constant effective area for the entire source extension, i.e., taken to be that of the source center, is not accurate enough if the source extension is large. In this case, \autoref{eqn:(4)} should be changed to 
\begin{flalign}
\label{eq:adjust01}
    &n^k_s=n_s\times\frac{\int t_k \int E_{\nu}^{-\gamma}\sum_{j} \omega_{j}A_{\rm eff}^{k}(E_{\nu}\,, \delta_j)dE_{\nu}}{\sum_{k^{'}} \int t_{k^{'}} \int E_{\nu}^{-\gamma}\sum_{j} \omega_{j}A_{\rm eff}^{k^{'}}(E_{\nu}, \delta_j)dE_{\nu}},
\end{flalign}
and \autoref{eqn:(6)} to
\begin{equation}
\label{eq:adjust02}
    S^{\rm ener}=\frac{\int E_{\nu}^{-\gamma}\sum_{j} \omega_{j}A_{\rm eff}^{k}(E_{\nu}, \delta_j)\mathcal{M}_{k}(E_{i}|E_{\nu}, \delta_j)dE_{\nu}}{\int E_{\nu}^{-\gamma}\sum_{j} \omega_{j}A_{\rm eff}^{k}(E_{\nu}, \delta_j)dE_{\nu}},
\end{equation}
where $\omega_j$ is the fraction of $S^{\rm spat}$ locating in the declination range $[\delta_j, \delta_{j+1})$. 

\section{The neutrino and gamma-ray connection}\label{sec:proton}

The spectral shape of neutrino emission is required to derive the upper limit on the hadronic gamma-ray flux. In p-p interactions, gamma-rays are produced accompanying the neutrino production. There is a clear connection of the spectrum and flux between these neutrinos and gamma-rays. Here we provide the way to derive the neutrino spectrum if the accompanying gamma-ray spectrum is known.
Similar to \cite{2020MNRAS.492.4246H}, we use the parameterized energy spectra of neutrinos and gamma-rays by \cite{2006PhRvD..74c4018K} to convert a certain hadronic gamma-ray spectrum into the neutrino spectrum in the hadronuclear (p-p) scenario.

The differential spectrum of neutrinos or gamma-rays $\Phi_{\nu,\gamma}(E)\equiv  d N_{\nu,\gamma}/d E$ can be given by  
\begin{flalign}
\label{eq:kelner02}
\begin{aligned}
\Phi_{\nu,\gamma}\left(E\right)
=\int_{E}^{\infty} \sigma_{\text{pp}}\left(E_{p}\right)\mathcal{G}_{p}\left(E_{p}\right) F_{\nu,\gamma}\left(\frac{E}{E_{p}}, E_{p}\right) \frac{d E_{p}}{E_{p}},
\end{aligned}
\end{flalign}
where
\begin{equation}
    \mathcal{G}_{p}(E_{p})\propto\int  n_{\rm H}\left(\vec{x}\right)J_{p}\left(E_{p},\vec{x}\right)dV,
\end{equation}
$\sigma_{\text{pp}}$ is the cross section of inelastic p-p interactions, $V$ is the volume of the emission region, $n_{\rm H}$ is the number density of the background gas, $J_{p}$ is the differential cosmic ray flux density, and $F_{\rm \nu}$ ($F_{\rm \gamma}$) is the energy distribution probability of neutrinos (gamma-rays) decayed from the secondary particles (e.g. $\rm \pi$-mesons and $\rm \eta$-mesons) in p-p interactions. 
We adopt the analytical parameterization of $F_{\gamma}$ and $F_{\nu}$ by \cite{2006PhRvD..74c4018K} (see Equations 58, 62, and 66 therein). The accuracy is better than 10\% in the cosmic ray energy range $100~{\rm GeV}\leq E_{p} \leq100~{\rm PeV}$ and for neutrino or gamma-ray energy of $E/E_{p}\geq10^{-3}$.

Given a gamma-ray spectrum $\Phi_\gamma(E)$, we use \autoref{eq:kelner02} to find out function $\mathcal{G}_{p}(E_{p})$, and hence derive the neutrino spectrum $\Phi_\nu(E)$ also with \autoref{eq:kelner02}. The p-p produced neutrino flavor ratio, after mixing in propagation, is $\nu_e:\nu_\mu:\nu_\tau\approx1:1:1$ at the Earth. So we calculate the muon neutrino flux $\Phi_{\nu_{\mu}+\bar{\nu}_{\mu}}$ as equal to one-third of the total all-flavor neutrino flux.

The gamma-ray spectrum is obtained by the best fit of the observational data. We assume a group of spectral models, as well as a pulsar wind model proposed by \citet{2003A&A...402..827A}. The group of spectral models are:
\begin{equation}\label{eq:best-fit}
\Phi_\gamma\propto\left\{
    \begin{aligned}
        & E_{10}^{-\gamma_{1}}  & & ({\rm PL})\\
        & E_{10}^{-\gamma_{1}-\gamma_{2}\,{\rm log_{10}}E_{10}} & & ({\rm LOGP})\\
        & E_{10}^{-\gamma_{1}}{\rm exp}\left(-E/E_{c}\right) & & ({\rm ECPL})\\
        & E_{10}^{-\gamma_{1}}{\rm exp}\left(-\sqrt{E/E_{c}}\right) & & ({\rm ECPL2})\\
    \end{aligned}
\right.
\end{equation}
where $E_{10}=E/10\,{\rm TeV}$. 
\autoref{table_para} shows the best-fit parameters for the gamma-ray spectra of the LHAASO sources, and the parameter values in the pulsar wind model\footnote{The dimensionless parameter $\mu$ describes the target density in the nebula, $\Gamma$ is the wind Lorentz factor, and $f_{p}$ is the fraction of spin down luminosity carried by protons.}. The source extension in \autoref{table_para} is taken according to either the measurements in TeV energy range or the prior extension of LHAASO sources (see details in \autoref{appendix:a}).

Note, in the spectral fitting, the gamma-ray absorption ($\gamma\gamma\rightarrow e^{+}e^{-}$) due to the interstellar radiation field (ISRF) and cosmic microwave background (CMB) is taken into account if the distances to the TeV counterparts of LHAASO sources are available in the \href{http://tevcat.uchicago.edu}{TeVCat} \citep{2008ICRC....3.1341W}. The ISRF energy density is taken from \cite{2017MNRAS.470.2539P}. The absorption is usually not important (see \autoref{table_tevcat}). 

All the neutrino spectra used for signal search are calculated through \autoref{eq:kelner02} except that the spectrum for the pulsar wind model is directly taken from \citet{2003A&A...402..827A}.
With the upper limit on neutrino flux, we can also get the upper limit on hadronic gamma-ray flux with a similar process.

We have tested in our calculation the accuracy of the solved-out function $\mathcal{G}_{p}$ by comparing the best-fit gamma-ray spectrum and the one derived by the solution function of $\mathcal{G}_{p}$. The difference is less than 2\% over an energy range $[E_{\gamma,1}, E_{\gamma,2}]$, which is corresponding to the neutrino energy range $[0.5E_{\gamma,1},0.5E_{\gamma,2}]$ where the central 90\%  neutrino events are located.

\begin{center}
\begin{table}
\centering
    \begin{subtable}{.9\columnwidth}
        \begin{tabular}{l c c c c c }
        \hline
        LHAASO & $\sigma_s$ & $\gamma_{1}$ & $\gamma_{2}$ & $E_{\rm c}$ & Model \\
         Source      & [deg] &             &              &   [TeV]   &   \\
        \hline
        J0534+2202 & 0.0  & 2.86 & 0.20 & --- & LOGP  \\
        J1825-1326 & 0.30 & 2.40 & 0.45 & --- & LOGP  \\
                   & 0.0  & 2.13 & ---  & 286 & ECPL  \\
        J1839-0545 & 0.41 & 2.26 & ---  & 36  & ECPL  \\
        J1843-0338 & 0.24 & 2.03 & ---  & 48  & ECPL  \\
        J1849-0003 & 0.09 & 1.99 & ---  & --- & SPL   \\
                   & 0.09 & 1.99 & ---  & 300 & ECPL  \\
        J1908+0621 & 0.52 & 2.53 & 0.30 & --- & LOGP  \\
        J2018+3651 & 0.20 & 1.57 & ---  & 26  & ECPL  \\
        J2032+4102 & 1.8  & 2.94 & ---  & --- & SPL   \\
        J2226+6057 & 0.24 & 2.29 & 0.33 & --- & LOGP  \\
        J1929+1745 & 0.30 & 2.40 & ---  & --- & SPL   \\
        J1956+2845 & 0.30 & 2.09 & ---  & --- & SPL   \\
        J2108+5157 & 0.0  & 1.95 & ---  & 20  & ECPL2 \\
        \hline
        \end{tabular}
    \end{subtable}
    \begin{subtable}{.9\columnwidth}
    \begin{tabular}{l c c c c c }
        \hline
        LHAASO  & $\sigma_s$ & $\mu$ & $\Gamma$ & $f_{p}$ & Model\\
        Source            & [deg]      &       &          &         &       \\
        \hline
        J0534+2202 & 0.0 & 1 & $10^{7}$ & 0.15 & Amato et al. \\
        \hline
    \end{tabular}
    \end{subtable}
\caption{The spectral parameters of hadronic gamma-rays and the source extensions ($\sigma_s$) used for signal search. The extensions refer to the source extensions of TeV gamma-ray counterparts given in \autoref{table_tevcat}.}\label{table_para}
\end{table}
\end{center}

\begin{figure*}
\begin{center}
\includegraphics[width=0.85\textwidth]{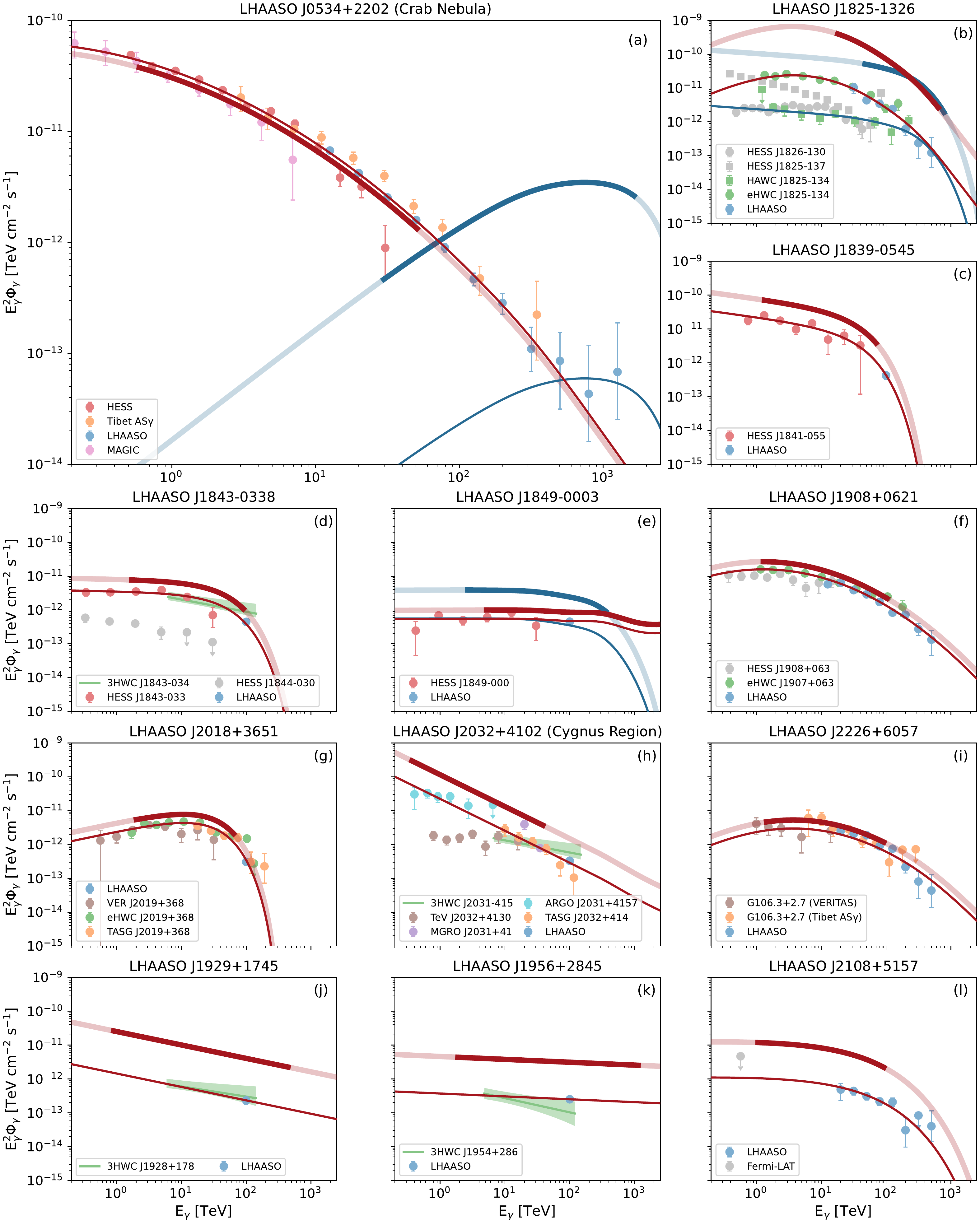}
\caption{The 90\% C.L. upper limits on hadronic gamma-ray flux (thick lines) from LHAASO sources in comparison with the observed gamma-ray flux (thin lines). The dark parts of the thick lines correspond to the central 90\% energy ranges in the neutrino detection. For each LHAASO source, the first model in \autoref{table_para} is shown in red, while the second model in blue. The blue data points are from LHAASO observations \citep{2021Natur.594...33C,425, 2021ApJ...919L..22C}. The other gamma-ray data are taken from various observations. The references for each LHAASO source are: (a) Crab Nebula by Tibet AS$\gamma$ \citep{2019PhRvL.123e1101A}, H.E.S.S. \citep{2006A&A...457..899A} and MAGIC \citep{2008ApJ...674.1037A}; (b) HESS J1826-130 \citep{2020A&A...644A.112H}, HESS J1825-137 \citep{2019A&A...621A.116H}, eHWC J1825-134 \citep{2020PhRvL.124b1102A}; (c) HESS J1841-055 \citep{2008A&A...477..353A}; (d) HESS J1843-033 and HESS J1844-030 \citep{2018A&A...612A...1H}, 3HWC J1843-034 \citep{2020ApJ...905...76A}; (e) HESS J1849-000 \citep{2018A&A...612A...1H}; (f) HESS J1908+063 \citep{2009A&A...499..723A}, eHWC J1907+063 \citep{2020PhRvL.124b1102A}; (g) VER J2019+368 \citep{2018ApJ...861..134A}, eHWC J2019+368 \citep{2020PhRvL.124b1102A}, TASG J2019+368 \citep{2021PhRvL.127c1102A}; (h) TeV J2032+4130 \citep{2014ApJ...783...16A}, MGRO J2031+41 \citep{2007ApJ...664L..91A, 2009ApJ...700L.127A}, 3HWC J2031-415 \citep{2020ApJ...905...76A}, TASG J2032+414 \citep{2021PhRvL.127c1102A}, ARGO J2031+4157 \citep{2014ApJ...790..152B}; (i) G106+2.7 by VERITAS \citep{2009ApJ...703L...6A} and Tibet $\rm{AS\gamma}$ \citep{2021NatAs...5..460T}; (j) 3HWC J1928+178 \citep{2020ApJ...905...76A}; (k) 3HWC J1954+286 \citep{2020ApJ...905...76A}; and (l) 4FGL J2108.0+5155 \citep{2021ApJ...919L..22C}. }\label{figure_bayes}
\end{center}
\end{figure*}

\section{Present constraints from IceCube data}\label{sec:result}

\begin{figure*}
\begin{center}
\includegraphics[width=0.86\textwidth]{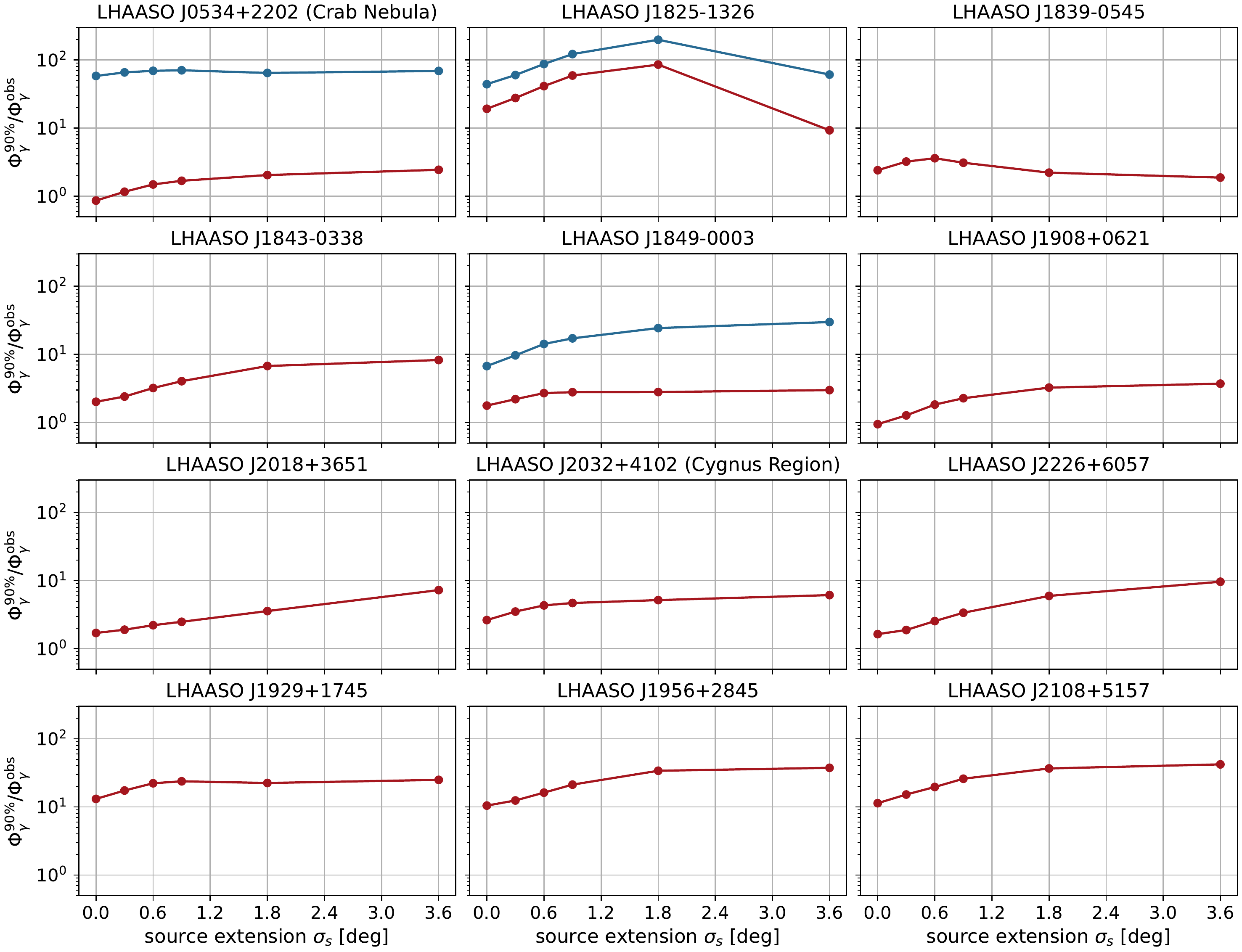}
\caption{The flux ratio between the 90\% C.L. upper limit on hadronic gamma-ray flux $\Phi^{90\%}_{\gamma}$ and the observed gamma-ray flux $\Phi^{\rm{obs}}_{\gamma}$ as function of source extension $\sigma_s$. The red (blue) lines correspond to the spectral shapes of hadronic gamma-rays following the red (blue) lines in \autoref{figure_bayes}. These results come from the analysis of the ten-year IceCube data.}\label{figure_extension}
\end{center}
\end{figure*}

Using the above-described method for the IceCube data, we derive the upper limits of hadronic gamma-ray flux from the LHAASO sources.
We compare the 90\% C.L. upper limits on the hadronic gamma-ray flux with the observed gamma-ray flux, as shown in \autoref{figure_bayes}. Only Crab Nebula is well constrained; its derived upper limit of hadronic gamma-ray flux disagrees with the best-fit observed gamma-ray flux within $5\sigma$ uncertainty, and is smaller than 86\% of the latter. This upper limit is somewhat higher than that by \cite{2022ApJ...925...85H}, becasue using \autoref{eq:kelner02} the additional gamma-rays from $\eta$-meson decay are considered. The IceCube data can only constrain the hadronic flux from the Crab Nebula below 50 TeV. However, although the central energy range covers the energy of hundreds of TeV in the cases of LHAASO J1825-1326, HAWC J1825-134, LHAASO J1849-0003, LHAASO J1929+1745, and LHAASO J1956+2845, their upper limits are higher than the gamma-ray fluxes observed, thus not constraining the hadronic component.


Some factors leading to uncertainties in the analysis should be noted here. The assumptions on the extension of neutrino sources affect the measurements of neutrinos/hadronic gamma-rays from LHAASO sources. \autoref{figure_extension} shows the dependence of the upper limit on the source extension. The extended-source hypothesis brings higher upper limits than those with the point-source hypothesis. The hadronic component can only be constrained in two cases under the point-like source assumption; besides Crab Nebula, the other is LHAASO J1908+0621, where the 90\% C.L. hadronic upper limit is lower than the gamma-ray flux within $\sim1\sigma$ uncertainty in the central 90\% energy range.
Moreover, the change of considering large extension (i.e., using \autoref{eq:adjust01} and \autoref{eq:adjust02}) is relatively large for the sources in the southern sky, for example, the expected number of signal events change from 5.4 to 5.7 for LHAASO J1839-0545 (the extension $\delta_s=0.9^{\circ}$), from 4.1 to 3.9 for LHAASO J1843-0338 ($\delta_s=0.9^{\circ}$), but for LHAASO J1908+0621 ($\delta_s=0.9^{\circ}$) only increases by 0.4\%.


The statistical method also affect the estimate of upper limits. We select 20 sources (including the seven sources associated with LHAASO sources) from Table III in the ten-year search paper 
\citep{2020PhRvL.124e1103A} and compare their upper limits\footnote{The upper limits for $E^{-2}$ ($E^{-3}$) spectrum are taken from Table III (Figure 3) of \cite{2020PhRvL.124e1103A}.} with the ones given by our Bayesian method. Our upper limits tend to be $1.4^{+0.6}_{-0.2}$ ($0.9^{+0.2}_{-0.3}$) times of the upper limits given by \cite{2020PhRvL.124e1103A} for the $E^{-2}$ ($E^{-3}$) neutrino spectrum.

The difference between the upper limits given by the two methods may be due to several reasons. The first is the difference in statistical approaches. The Frequentist approach is used in the ten-year search, while we use the Bayesian approach. The results given by Bayesian approach relies on the uniform prior distribution $\pi(n_s)$. The second is the difference in the likelihood functions. The likelihood function used in the ten-year search includes two free parameters ($n_s$ and $\gamma$), while our likelihood function only one ($n_s$). The third is the difference in the background PDFs. The spatial and energy background PDFs ($B^{\rm spat}$ and $B^{\rm ener}$) are precisely parameterized with good data/Monte Carlo agreement \citep{2016ApJ...833....3A,2019EPJC...79..234A} 
while ours are obtained from the data. 

Furthermore, we also estimate the significance of signal events as function of the unknown source extension (see \autoref{appendix:b} for detailed calculation). We find that three LHAASO sources show relatively strong significance of signal neutrino events for some extension values: LHAASO J2032+4102 if $\delta_s=0.3^{\circ}$; LHAASO J1929+1745 if $\delta_s=0.6^{\circ}$; and LHAASO J1908+0621 under PS assumption.

\begin{figure*}
\begin{center}
\includegraphics[width=0.86\textwidth]{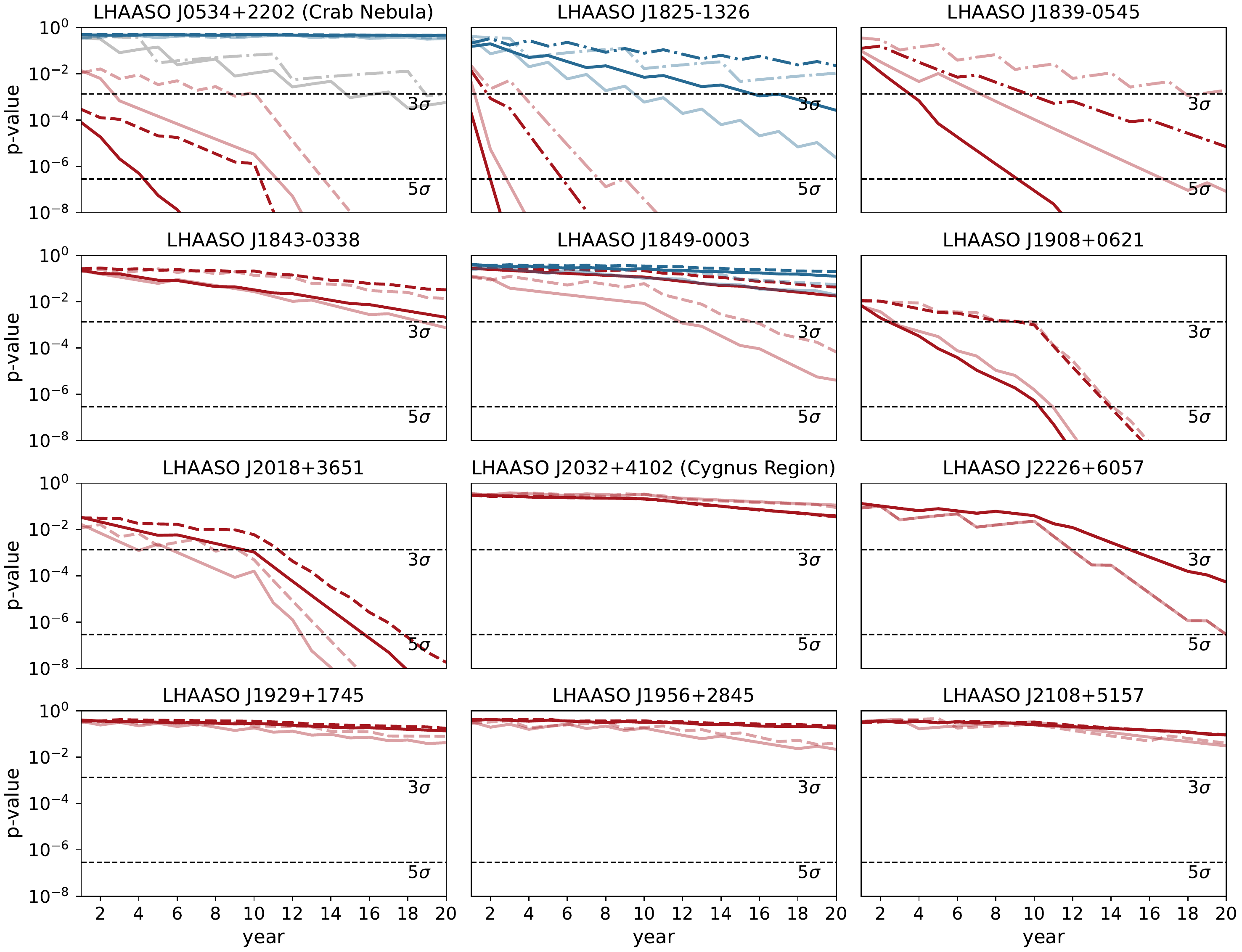}
\caption{The statistical significance (p-value) of the future track-event observation as function of observation time. The starting point (zero time) is the present status, already with the 10 year (2008-2018) IceCube data. The red and blue lines correspond to the spectral models in red and blue in \autoref{figure_bayes}, respectively. The dark and light lines represent the observation with low and high threshold energy, $E_{\rm th}=1$ and $10\,{\rm TeV}$, respectively. In the first panel for Crab Nebula, the gray lines present the pulsar wind model of \citet{2003A&A...402..827A} but with $E_{\rm th}=100\,{\rm TeV}$: the gray solid line is for the case of 20 year observation by a detector 30 times larger than IceCube at the South Pole; and the gray dashed-dotted line is for 20 year observation by a detecor 30 times larger than IceCube at the position of Lake Baikal. As for the red and blue lines, the solid lines present the cases of the 10 year IceCube observation followed by 10 year $\rm{PLE\nu M\hbox{-}1}$ and then 10 year $\rm{PLE\nu M\hbox{-}2}$ observation; the dashed lines present the cases for the sources at $\delta_s>-5^{\circ}$, the 10 year IceCube observation followed by 10 year more IceCube and then 10 year IceCube-Gen2 observation; and the dashed-dotted lines are the cases of the sources at $\delta_s<-5^{\circ}$ observed by a IceCube-like detector at the position of Lake Baikal for 20 years.}\label{figure_pvalue}
\end{center}
\end{figure*}


\section{Future Prospects}\label{sec:prospects}

In this section we evaluate the detection of neutrinos from LHAASO sources by current and future neutrino telescopes, assuming that all the observed gamma-ray emission is hadronic origin. In particular, we estimate the observational time it takes before obtaining a $5\sigma$ detection of the source. We consider the Planetary Neutrino Monitoring System ($\rm{PLE\nu M}$), proposed by \cite{2021arXiv210713534S}, a concept of global repository of high-energy neutrino observations by current and future neutrino telescopes. Two proposed systems are considered: $\rm{PLE\nu M\hbox{-}1}$ consisting of IceCube and three IceCube-like detectors, located at the latitudes of Baikal-GVD, KM3NeT-ARCA and P-ONE; and $\rm{PLE\nu M\hbox{-}2}$ consisting of IceCube-Gen2 and the three IceCube-like detectors as those in $\rm{PLE\nu M\hbox{-}1}$. Here by IceCube-like we mean the effective area is similar to IceCube, and we assume IceCube-Gen2 has an effective area 7.5 times that of IceCube. 

To quantitatively describe the future observation results, we estimate the statistical significance of observation with a p-value analytically expressed as \citep{CMS-NOTE-2011-005, 2017APh....86...46H}
\begin{flalign}
p_{\text {value}}&=\frac{1}{2}\left[1-\operatorname{erf}\left(\sqrt{q_{0}^{\rm obs} / 2}\right)\right],
\end{flalign}
where
\begin{flalign}
q_{0}^{\rm obs}&=2\left[Y_{b}-N_{D}+N_{D} \ln \left(\frac{N_{D}}{Y_{b}}\right)\right],
\end{flalign}
$Y_{b}$ is the expected number of background events, and $N_{D}$ is the median of Poisson-distributed events containing both signal and background. We count events within a solid angle $\Omega=\pi\sigma_{\rm eff}^2$ around the source, where $\sigma_{\rm eff}=\sqrt{\sigma_{s}^2+(1.6\Delta\xi_{\rm det})^{2}}$. The angular resolution $\Delta\xi_{\rm det}$ is around $0.4^{\circ}$ for IceCube and $0.2^{\circ}$ or smaller for the detectors in water, e.g., Baikal-GVD, KM3NeT and P-ONE. The angle $\Omega$ is the one that contains roughly 72\% of signal events from the source. This method has been applied by \cite{2017APh....86...46H} to evaluating the detection of the neutrino emission from Galactic sources with IceCube. 

When we calculate the number of detected neutrino events, we take the following approach. First, only neutrinos with energy above some detection threshold $E_{\rm th}$ are counted.
Second, as the background events in up-going track events are mainly induced by atmospheric muon neutrinos, the software MCEq \citep{2015EPJWC..9908001F} is employed to model the background from atmospheric neutrinos, assuming the Gaisser-H4a model \citep{2012APh....35..801G} for primary cosmic rays and the Sibyll2.3c model \citep{2019PhRvD.100j3018F} for hadronic interactions. Third, only the up-going track events with the zenith angle $\theta_z>85^{\circ}$ are considered, because the discovery potential for IceCube is much larger in the northern sky than in the south,  by a few or a few tens \citep[see Figure 3 of][]{2020PhRvL.124e1103A}. Fourth, we consider that a certain source is only visible if $\theta_z>85^{\circ}$, and the visible time is calculated taking into account the Earth's rotation.


The results about the prediction of the significance of future observations on the 12 LHAASO sources are presented in \autoref{figure_pvalue}. In the calculation, we assume that the hadronic gamma-rays follow the energy spectra in \autoref{figure_bayes}. We also assume the situation of future observations are: following the 10-year (2008-2018) observation by IceCube is the operation of $\rm{PLE\nu M\hbox{-}1}$ for 10 years, and then $\rm{PLE\nu M\hbox{-}2}$ for the next 10 years.

We see that within 20 years, five LHAASO sources can be discovered at the level of $5\sigma$ if the threshold is $E_{\rm th}=1\,{\rm TeV}$: Crab Nebula, LHAASO J1825-1326, LHAASO J1839-0545, LHAASO J1908+0621, and LHAASO J2018+3651; one source can also be detected at the level of $5\sigma$ but for $E_{\rm th}=10\,{\rm TeV}$: LHAASO J2226+6057; and two sources are possible to be detected with $3\sigma$ C. L. within 20 years: HAWC J1825-134 and LHAASO J1849-0003 (SPL model).

Consider the neutrino search only by IceCube and IceCube-Gen2 (shown as the dashed lines in \autoref{figure_pvalue}), 4 sources can be discovered within 20 years, i.e., Crab Nebula, LHAASO J1908+0621, LHAASO J2018+3651, and LHAASO J2226+6057. The first three are consistent with the predictions in Figure 18 of \cite{2021JPhG...48f0501A}. 

For the combined search by $\rm{PLE\nu M\hbox{-}1}$ and $\rm{PLE\nu M\hbox{-}2}$, the IceCube-like detectors located at the northern hemisphere is very helpful in observing sources around the equatorial plane or in the southern sky ($\delta_s<-5^{\circ}$). For an example of LHAASO J1825-1326, if all the gamma-rays are hadronic in origin then even one IceCube-like detector at the latitude of Baikal-GVD can provide a $5\sigma$ detection within about 6 years.

As for Crabe Nebula, if consider only the neutrino emission in the pulsar wind model proposed by \citet{2003A&A...402..827A}, the signals are hard to be discovered. In order to observe at the level of $3\sigma$ within 20 years, a large neutrino detector with the effective area 30 times larger than IceCube is required (see the gray lines in \autoref{figure_pvalue}).

\section{Summary and discussion}\label{sec:summary}

We analyze the ten-year IceCube data with Bayesian approach to constrain the hadronic gamma-ray flux from the directions of LHAASO sources, and give the 90\% C.L. upper limits for both extended- and point-source hypotheses. The results given by the Bayesian approach rely on the uniform prior distribution. We further evaluate the combined search for high energy neutrinos from LHAASO sources by IceCube, IceCube-Gen2 and three IceCube-like detectors located at the latitudes of Baikal-GVD, KM3NeT-ARCA, and P-ONE, respectively.

To summarize, the main conclusions are: \textbf{i)} The 90\% C.L. upper limit on hadronic gamma-ray flux from Crab Nebula is 86\% of total gamma-ray flux observed, disfavoring hadronic origin of its gamma-rays; However, no constraint can be made for gamma-ray radiation mechanism in the other LHAASO sources by the present IceCube data; \textbf{ii)} Six LHAASO sources are expected to be discovered at the level of $5\sigma$ by $\rm{PLE\nu M\hbox{-}1}$ and $\rm{PLE\nu M\hbox{-}2}$ within 20 years, if their gamma-rays are totally hadronic in origin; \textbf{iii)} An IceCube-like detector at the location of Baikal-GVD is expected to give a $5\sigma$ discovery around 6 years if all the gamma-rays from LHAASO J1825-1326 are hadronic in origin.

In our analysis, the three most significant sources in neutrino searches are  LHAASO J2032+4102, LHAASO J1929+1745 and LHAASO J1908+0621. Under the point-source hypothesis, the pre-trial p-value is 0.051 for LHAASO J1908+0621 and 0.091 for LHAAS0 J2032+4102 which are similar with the p-values in the directions of MGRO J1908+06 and 2HWC J2031+415 in the ten-year search by IceCube \citep{2020PhRvL.124e1103A}. The TeV counterparts of LHAASO J1929+1745 are not in the source list of ten-year search because they are not bright in the \href{http://tevcat.uchicago.edu}{TeVCat} \citep{2008ICRC....3.1341W}. LHAASO J1929+1745 is even one of the faintest LHAASO sources at 100 TeV \citep{2021Natur.594...33C}. \cite{2020ApJ...897..129M}  tried to explain the gamma-rays from 2HWC J1928+177 ($0.10^{\circ}$ from LHAASO J1929+1745) with a hadronic accelerator model. 


The neutrino constraint on the Crab Nebula disfavors that hadronic process dominates the bulk of the observed gamma-ray emission. However, the gamma-ray spectrum by LHAASO shows a possible hardening around PeV energies \citep{425}. If the PeV gamma-rays are hadronic in origin, a neutrino detector with the effective area $\sim30$ times larger than IceCube is required to observe the neutrinos associated with these PeV gamma-rays at the level of $3\sigma$ within 20 years (see the first panel in \autoref{figure_pvalue}). 

Baikal-GVD is with 8 clusters deployed and two more clusters per year are planned from 2022 to 2024. It will occupy a water volume around $0.7\,{\rm km^3}$ in total \citep{2021arXiv210606288B}. A KM3NeT-ARCA block comprises 115 strings and eleven detection units have been deployed \citep{KM3NET}. P-ONE has two pathfinders deployed and the design is underway \citep{2021EPJC...81.1071B}. The P-ONE Explorer (10 strings) is planned to be deployed in 2023-2024 and remainder of the array (70 strings) is planned for deployment between 2028-2030 \citep{2020NatAs...4..913A}. IceCube-Gen2 is planned for deployment between 2027-2033. Thus $\rm{PLE\nu M\hbox{-}1}$ will start to operate after 2030, while $\rm{PLE\nu M\hbox{-}2}$ after 2033. The neutrino source candidates are expected to be identified or excluded at the high confidence level from 2030 to 2050.

\section*{Acknowlegdments}
We thank Hao Zhou and Xiao-Yuan Huang for useful discussions. This work is supported by the Natural Science Foundation of China (No. 11773003, U1931201) and the China Manned Space Project (CMS-CSST-2021-B11).

\section*{Data Availability}
The ten-year (2008-2018) muon-track data by IceCube are available at \href{https://icecube.wisc.edu/data-releases/2021/01/all-sky-point-source-icecube-data-years-2008-2018/}{https://icecube.wisc.edu/data-releases/2021/01/all-sky-point-source-icecube-data-years-2008-2018/}. The TeVCat online source catalog is available at \href{http://tevcat.uchicago.edu}{http://tevcat.uchicago.edu}. The ISRF energy density is available at \href{https://cdsarc.cds.unistra.fr/viz-bin/cat/J/MNRAS/470/2539}{https://cdsarc.cds.unistra.fr/viz-bin/cat/J/MNRAS/470/2539}. The code MCEq is available at \href{https://github.com/afedynitch/MCEq}{https://github.com/afedynitch/MCEq}. Other data are available through the references in the captions of \autoref{figure_bayes} and \autoref{table_tevcat}.


\bibliographystyle{mnras}
\bibliography{example}



\appendix

\begin{appendices}

\section{TeV gamma-ray Counterparts}\label{appendix:a}

In this appendix, we describe in some details about the TeV counterparts of LHAASO sources. \autoref{table_tevcat} gives the distances and the extensions of these TeV counterparts.

\begin{table*}
\begin{center}
\begin{tabular*}{0.87\textwidth}{l c l c c c c c}
\hline
\hline
LHAASO source & Extension &TeV Counterpart  & Distance  & $\rm exp(-\tau)$ &Angular Distance & Extension & Reference \\
              &   [deg]   &                 & [kpc]     & & [deg]            & [deg] &\\
\hline
J0534+2202 & \textbf{PS}   & Crab Nebula    &2.0  &0.99 &0.08 & 0.01 & [1] \\
\hline
J1825-1326 & \textbf{0.30} & HESS J1826-130 &4.0  &0.92 &0.39  & 0.21 & [2] \\
                  &     &HESS J1825-137  &3.9 &0.92 &0.39  & 0.55 (0.51) & [3] \\
                  &      &eHWC J1825-134  & --- &--- & 0.09 & 0.36       & [4] \\
                  &      &HAWC J1825-134  & --- &--- & 0.03 & \textbf{PS}         & [5] \\
\hline
J1839-0545 & ES   & HESS J1841-055  &--- &--- &0.29 & \textbf{0.41}     & [6] \\
\hline
J1843-0338 & ES   &HESS J1843-033  &--- &--- &0.22 & \textbf{0.24}         & [6] \\ 
                  &      &HESS J1844-030  &---&---  &0.69 & PS           & [6] \\
                  &      &3HWC J1843-034  &---&---  &0.30 & PS           & [7] \\
\hline
J1849-0003 & ES   &HESS J1849-000  &7 &0.84   &0.11 & \textbf{0.09}  & [6] \\
\hline
J1908+0621 & 0.58 &HESS J1908+063  &---&--- &0.11 & 0.34         & [8]\\ 
                  &      &eHWC J1907+063  &---&--- &0.14 & \textbf{0.52}         & [4]\\ 
\hline
J2018+3651 & ES    &VER J2019+368  &---&--- &0.10 & 0.34 (0.14)   & [9] \\ 
                  &       &eHWC J2019+368 &---&--- & 0.17& \textbf{0.20}         & [4] \\ 
                  &       &TASG J2019+368 &---&--- & 0.19& 0.28         & [10] \\ 
\hline
J2032+4102 & ES    & 3HWC J2031+415  &---&---  &0.47 &PS         & [7]\\ 
                  &       & TeV J2032+4130  & 1.8 &0.98 &0.52 &0.16 (0.07) & [11]\\ 
                  &       & ARGO J2031+4157 & 1.4 &0.99 &1.5  &\textbf{1.8}       & [12] \\ 
                  &       & TASG J2032+414  & --- &--- &0.41 & PS        & [13] \\ 
                  &       & MGRO J2031+41   & ---& &0.51 & 3.0       & [14] \\
\hline
J2226+6057 & 0.36  & G106.3+2.7 (VERITAS) & 0.8  &0.99 &0.14 &0.27 (0.18)  & [15]  \\ 
                  &       & G106.3+2.7 (${\rm Tibet\,AS_\gamma}$) & 0.8 &0.99 &0.11 & \textbf{0.24}  & [16] \\
\hline
J1929+1745 & \textbf{ES}    & 3HWC J1928+178 & --- &--- & 0.16 & PS  & [7]\\
\hline
J1956+2845 & \textbf{ES}    & 3HWC J1954+286 & --- &--- & 0.33 & PS & [7]\\
\hline
J2108+5157 & \textbf{PS}  & --- & --- & ---& ---  & --- & --- \\
\specialrule{0em}{1pt}{1pt}
\hline
\end{tabular*}
\caption{The TeV gamma-ray counterparts associated with the LHAASO sources. The second column is the intrinsic extension from LHAASO measurements. PS represents point-like source whereas ES extended source. The prior intrinsic extension for extended LHAASO source is $0.3^{\circ}$ \citep{2021Natur.594...33C}. The seventh column is the intrinsic extension of TeV counterparts. The number inside (outside) the brackets is the extension along the minor (major) axis. The extension in \textbf{boldface} indicates the source extension adopted in \autoref{table_para}. The distance from the TeV counterpart to us is taken from \href{http://tevcat.uchicago.edu}{TeVCat} \citep{2008ICRC....3.1341W}. The flux attenuation $\rm exp(-\tau)$ due to the ISRF and CMB is given at $E_{\gamma}=100\,{\rm TeV}$. The extensions of TeV counterparts and their angular distances to the LHAASO source refer to [1] \protect\cite{2020NatAs...4..167H}, [2] \protect\cite{2020A&A...644A.112H}, [3] \protect\cite{2019A&A...621A.116H}, [4] \protect\cite{2020PhRvL.124b1102A}, [5] \protect\cite{2021ApJ...907L..30A}, [6] \protect\cite{2018A&A...612A...1H}, [7] \protect\cite{2020ApJ...905...76A}, [8] \protect\cite{2009A&A...499..723A}, [9] \protect\cite{2018ApJ...861..134A}, [10] \protect\cite{2021PhRvL.127c1102A}, [11] \protect\cite{2014ApJ...783...16A}, [12] \protect\cite{2014ApJ...790..152B}, [13] \protect\cite{2021PhRvL.127c1102A}, [14] \protect\cite{2007ApJ...664L..91A}, [15] \protect\cite{2009ApJ...703L...6A} and [16] \protect\cite{2021NatAs...5..460T}. }\label{table_tevcat}
\end{center}
\end{table*}

\textbf{Crab Nebula.} LHAASO-KM2A has detected an 1.1 PeV photon from the direction of Crab Nebula and its energy spectrum has a possible hardening around PeV energies which indicates a hadronic component \citep{425}. So we consider the model by \citep{2003A&A...402..827A}, in which protons take $f_p=60\%$ of pulsar wind energy and produce gamma-rays and neutrinos through p-p interactions. The target density $n_t$ is expressed as
\begin{equation}
    n_{t}=10\mu(M_{N\odot}/R_{\rm pc}^3)\,{\rm cm}^{-3}
\end{equation}
with the parameter $\mu$ defined in the Equation 9 of \cite{2003A&A...402..827A}. The p-p interactions are dominant in comparison with the photo-meson ($p\gamma$) productions if $\mu$ is not very much smaller than unity. 
As for the target density $\mu=1$, the gamma-ray flux will be higher than the observed flux if $f_p>15\%$ and the peak energy will be much lower than PeV if the wind Lorentz factor $\Gamma<10^{7}$. The density parameter $\mu$ equals to 5 if the mass estimated by \cite{1997AJ....113..354F} uniformly distributes in the nebula \citep{2003A&A...402..827A}. The higher target density requires lower $f_p$ to explain the observation.

\textbf{LHAASO J1825-1326.} HAWC has resolved the region around eHWC J1825-134 into three sources (HAWC J1825-138, HAWC J1826-128 and HAWC J1825-134) and discovered the point-like source HAWC J1825-134 whose energy spectrum extends well beyond 200 TeV without a cutoff \citep{2021ApJ...907L..30A}. As the angular distance between HAWC J1825-134 and LHAASO J1825-1326 is only $0.03^{\circ}$ (see \autoref{table_tevcat}), we fit the energy spectrum of HAWC J1825-134 in combination of the measurements by HAWC and LHAASO. The extrapolated flux ($>100\,{\rm TeV}$) of HAWC J1825-138 and HAWC J1826-128 are removed from the measurements by LHAASO.

\textbf{LHAASO J2032+4102.} The angular distances between LHAASO J2032+4102 and its TeV counterparts are larger than the angular resolution of LHAASO-KM2A around 15-20 arcmin at 100 TeV \citep{2021Natur.594...33C}. Such angular distances indicate that hundred TeV gamma-rays and TeV gamma-rays are probably generated in different emission regions or astrophysical sources. Further studies on the morphology and spectrum of this region are required to reveal the origin of PeV gamma-rays from Cygnus region.

\textbf{LHAASO J2108+5157.} LHAASO J2108+5157 is the only LHAASO source without TeV counterparts. As the constraint on TeV gamma-ray flux is not that strong, we refer the hadronic model shown in Figure 5 of \cite{2021ApJ...919L..22C} and use the ECPL2 in \autoref{eq:best-fit} to model the profile of hadronic gamma-ray spectrum.

\textbf{Rescaling due to extension.} In \autoref{figure_bayes}, the flux measurements of HESS J1826-130, HESS J1841-055, HESS J1908+063, VER J2019+368, TeV J2032+4130 and G106.3+2.7 (VERITAS) are rescaled according to their intrinsic extensions and integration regions. See details in Appendix B of \cite{2022ApJ...925...85H}. 

\begin{figure*}
\begin{center}
\includegraphics[width=0.86\textwidth]{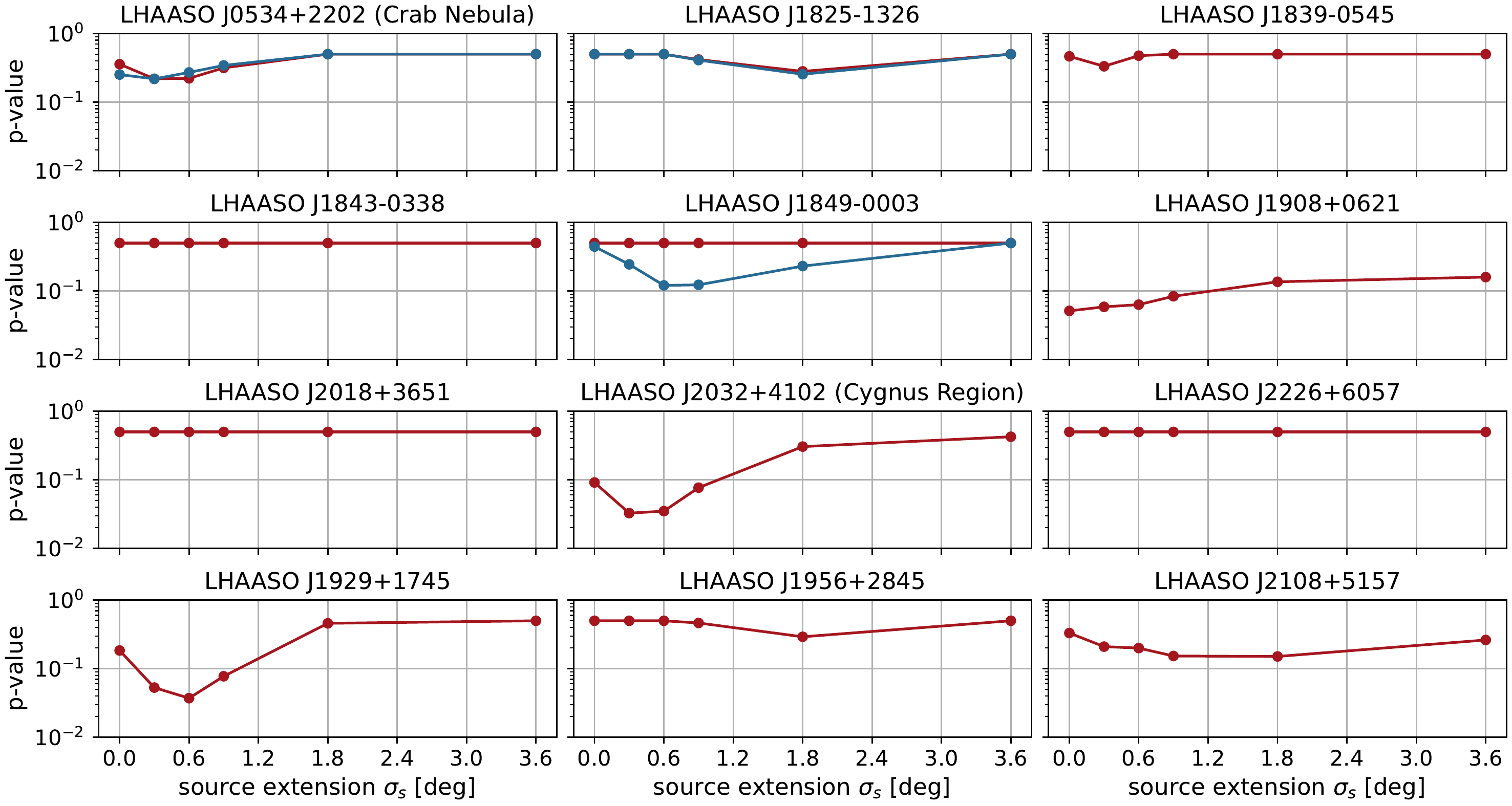}
\caption{The statistical significance (p-value) of neutrinos from LHAASO sources as function of source extension $\sigma_s$.  The red (blue) lines correspond to the spectral models in red (blue) lines in \autoref{figure_bayes}. These results are from the ten-year IceCube data.}\label{figure_pvalue_extension}
\end{center}
\end{figure*}

\section{Signal Significance for Different Source Extensions}\label{appendix:b}

In addition to the upper limits shown in \autoref{figure_bayes} and \autoref{figure_extension}, we give the significance of signal neutrino events from the direction of LHAASO sources here. A test statistic ($TS$) is built as 
\begin{equation}
    {TS}=2{\rm ln}\left[\frac{L\left(X\big|\hat{n}_s, \sigma_s, \Phi^{\rm model}_{s}\right)}{L\left(X\big|n_s=0, \sigma_s, \Phi^{\rm model}_{s}\right)}\right],
\end{equation}
where $L$ is the likelihood function in \autoref{eq:likelihood}, $X$ represents the observed data and $\hat{n}_s$ is the number of signal events ($n_s\geq0$) maximizing the likelihood with given source extension $\sigma_s$ and spectral shape $\Phi^{\rm model}_{s}$ (e.g. $E^{-\gamma}$) of incident astrophysical neutrinos. Assuming the validity of the Wald approximation \citep{10.2307/1990256}, $TS$ will follow the PDF as
\begin{equation}
    f({TS})=\frac{1}{2}\delta\left({ TS}\right)+\frac{1}{2}\frac{1}{\sqrt{2\pi}}\frac{1}{\sqrt{TS}}e^{-{TS}/2}
\end{equation}
for pure background hypothesis \citep{2011EPJC...71.1554C}. So the statistical significance of signal neutrino events can be described with the pre-trial p-value as $1-F(TS)$, where $F(TS)$ is the CDF of $f(TS)$:
\begin{equation}
    F(TS)=\frac{1}{2}\left[1+\operatorname{erf}\left(\sqrt{TS/2}\right)\right].
\end{equation}
The p-values for different source extensions are shown in \autoref{figure_pvalue_extension}.

\end{appendices}

\bsp	
\label{lastpage}
\end{document}